\documentclass[10pt,a4paper]{article}
\usepackage{amsmath,amssymb,epsfig,cite,bm}

\usepackage[]{graphicx,amsmath,amssymb,bm}

\title{The radiative self-force and charged fluids} 

\author{David A. Burton\thanks{Physics Department, Lancaster University, Lancaster, UK and Cockcroft Institute, Daresbury, Warrington, UK}
\and
Anthony Carr\footnotemark[1]
\and
Jonathan Gratus\footnotemark[1]
\and
Adam Noble\thanks{Department of Physics, SUPA, University of Strathclyde, Glasgow, UK}
}
\begin{document} 
\maketitle 
\begin{abstract}
We develop a new fluid model of a warm plasma that includes the radiative self-force on each plasma electron. Our approach is a natural generalization of established methods for generating fluid models without radiation reaction. The equilibrium of a magnetized plasma is analysed, and it is shown that the thermal motion is confined to the magnetic field lines. A dispersion relation is deduced for electric waves in a magnetized plasma, and it is shown to agree with our recently established relativistic kinetic theory derived from the Lorentz-Abraham-Dirac equation. 
\end{abstract}

\section{Introduction}
Contemporary advances in ultra-intense laser facilities have driven the recent surge of interest in the collective behaviour of charged matter in extreme conditions, and a particularly vexing topic in that context concerns the coupling of an electron to its own radiation field. An accelerating electron emits electromagnetic radiation, and the energy and momentum carried away by the electromagnetic field must be accounted for. In most practical cases, the Lorentz force on an electron, due to an applied electromagnetic field, is considerably larger than the force due to the electron's emission and the effect of the recoil due to the emitted radiation is negligible or can be adequately represented using simple physical reasoning.
Although such arguments avoid the difficulties that plague more comprehensive analyses, the parameter regimes promised by forthcoming ultra-intense laser facilities ensure that more fundamental considerations are now of practical necessity. For example, ELI~\cite{eli} is expected to operate with intensities $10^{23} {\rm W}/{\rm cm}^2$ and electron energies in the ${\rm GeV}$ range, at which level the radiation reaction force becomes comparable to, and can even exceed, the applied force due to the laser field.

One of the most notorious equations in physics was developed by Dirac~\cite{dirac:1938} to describe a classical point electron's radiative self-force. The Lorentz-Abraham-Dirac (LAD) equation is a fully relativistic description of a structureless point particle in an applied electromagnetic field $F_{ab}$ and has the form
\begin{equation}
\label{LAD}
\frac{d^2 x^a}{d\lambda^2} = -\frac{q}{m} F^a{ }_b\,\frac{dx^b}{d\lambda} + \tau \Delta^a{ }_b \frac{d^3 x^b}{d\lambda^3}
\end{equation}
with $q$ the particle's charge, $m$ the particle's rest mass, $\tau = q^2/6\pi m$ in Heaviside-Lorentz units with $c=\epsilon_0=\mu_0=1$, and the tensor $\Delta^a{ }_b$ is
\begin{equation}
\Delta^a{ }_b = \delta^a_b + \frac{dx^a}{d\lambda} \frac{dx_b}{d\lambda}.
\end{equation}
For an electron, $q=-e<0$.
The Einstein summation convention is used throughout the present article, indices are raised and lowered using the metric tensor $\eta_{ab} = {\rm diag}(-1,1,1,1)$ and lowercase Latin indices range over $0,1,2,3$. The particle's $4$-velocity $dx^a/d\lambda$ is normalized as follows:
\begin{equation}
\label{proper_time}
\frac{dx^a}{d\lambda} \frac{dx_a}{d\lambda} = -1
\end{equation}
where $\lambda$ is the particle's proper time.

Dirac derived (\ref{LAD}) for an electron by appealing to the conservation condition on the stress-energy-momentum tensor (see Ref. [\citenum{ferris:2011}] for a recent discussion of the derivation). Dirac's approach requires a regularization of the electron's singular contribution to the stress-energy-momentum tensor followed by a renormalization of the electron's rest mass. The procedure leads to the third-order term in (\ref{LAD}), which is the source of the famous pathological behaviour exhibited by solutions to the LAD equation (see Ref. [\citenum{hammond:2010}] for a recent discussion).

The standard approach to ameliorating the problems with the LAD equation is to replace the third-order terms in (\ref{LAD}) (radiation reaction force) with the derivative of the first term on the right-hand side of (\ref{LAD}) (the applied Lorentz force). This procedure is justifiable if the radiation reaction force is a small perturbation to the Lorentz force, and it yields the Landau-Lifshitz (LL) equation:
\begin{equation}
\label{LL}
\frac{d^2 x^a}{d\lambda^2} = -\frac{q}{m} F^a{ }_b\,\frac{dx^b}{d\lambda} - \tau\frac{q}{m} \partial_c F^a{ }_b \frac{dx^b}{d\lambda} \frac{dx^c}{d\lambda}+ \tau \frac{q^2}{m^2} \Delta^a{ }_b F^b{ }_c F^c{ }_d \frac{dx^d}{d\lambda}.
\end{equation}
Unlike the LAD equation, the LL equation is second order in derivatives in $\lambda$ and its solutions are free from pathologies.

Recent years have seen a substantial growth of interest in kinetic theories incorporating radiation reaction (see, for example, Refs. [\citenum{tamburini:2011},~\citenum{lehmann:2012}]), and almost all such theories are based on the LL equation from the outset. However, the briefest of glances at (\ref{LAD}) and (\ref{LL}) suggests that a degree of mathematical clarity (with concomitant physical insight) is likely to be gained by starting with a kinetic theory based on the LAD equation. Of course, procedures for extracting physically acceptable behaviour from the LAD kinetic theory must be introduced during the analysis. 

Remarkably, until recently~\cite{noble:2013}, very little has appeared in the literature concerning a fully relativistic many-body system governed by the LAD equation. In addition, some of what can be found~\cite{hakim:1968} appears to lead to physically inconsistent results (see Ref.[\citenum{noble:2013}] for a discussion of this point). Perhaps one of the reasons why the LAD equation appears to have been neglected in the physics literature is the necessary, but unconventional, introduction of the notion of a `phase' space encoding acceleration as well as velocity and spacetime events. By employing the appropriate geometrical machinery, we recently showed~\cite{noble:2013} how to construct a Vlasov equation based on the LAD equation.

The purpose of the present article is to demonstrate how the Vlasov equation developed in Ref. [\citenum{noble:2013}] naturally induces a warm fluid theory and to explore the consequences of that theory. Although we followed a similar approach in Ref. [\citenum{noble:2011}], the warm fluid approximation used there did not treat the acceleration and velocity moments on an equal footing. Our aim in the following is to redress that balance and obtain a more general warm fluid theory. We also use our new theory to explore electric waves in a magnetized plasma.
\section{Kinetic theory}
As shown in Ref. [\citenum{noble:2013}], the LAD equation may be written in first-order form as
\begin{align}
&\frac{d x^a}{d\lambda} =  \dot{x}^a,\\
&\frac{d v^\mu}{d\lambda} = a^\mu,\\
&\frac{d a^\mu}{d\lambda} = \ddot{x}^a \ddot{x}_a v^\mu + \frac{1}{\tau}\bigg (a^\mu + \frac{q}{m} F^\mu{ }_a \dot{x}^a\bigg)
\end{align}
in terms of the proper-velocity (celerity) $\bm{v} = (v^1,v^2,v^3)$ and acceleration $\bm{a}=(a^1,a^2,a^3)$ in a $10$-dimensional `phase' space $(x^a, v^\mu, a^\nu)$ where lowercase Greek indices range over $1,2,3$. The shorthand $\dot{x}^0 = \sqrt{1+\bm{v}^2}$, $\ddot{x}^0 = a^\mu v_\mu / \sqrt{1+\bm{v}^2}$, $\dot{x}^\mu = v^\mu$, $\ddot{x}^\mu = a^\mu$ have been used, and the parametrization in terms of $\bm{v}$ and $\bm{a}$ has been chosen to satisfy the constraints $\dot{x}^a \dot{x}_a = -1$, $\ddot{x}^a \dot{x}_a = 0$ arising from (\ref{proper_time}) and its first derivative with respect to $\lambda$. Greek indices are raised and lowered using the Kronecker delta $\delta_{\mu\nu}$. 
 
From a geometrical perspective, the Vlasov equation may be understood as the preservation of a differential form of maximal degree along the flow of single-particle orbits. For present purposes, the $1$-particle distribution $f= f(x,\bm{v},\bm{a})$ of plasma electrons may be naturally extracted from that differential form~\cite{noble:2013} leading to 
\begin{equation}
\label{generalized_vlasov}
Lf + \frac{3}{\tau} f = 0 
\end{equation}
where $L$ is the Liouville operator
\begin{equation}
L = \dot{x}^a \frac{\partial}{\partial x^a} + a^\mu \frac{\partial}{\partial v^\mu} + \bigg[\ddot{x}^a \ddot{x}_a v^\mu + \frac{1}{\tau}\bigg (a^\mu + \frac{q}{m} F^\mu{ }_a \dot{x}^a\bigg) \bigg]\frac{\partial}{\partial a^\mu}.
\end{equation} 
The second term on the left-hand side of (\ref{generalized_vlasov}) may be understood as a consequence of losses due to radiation (for more details see Ref. [\citenum{noble:2013}]).

Maxwell's equations are
\begin{align}
\label{cov_faraday}
&\partial_a F_{bc} + \partial_c F_{ab} + \partial_b F_{ca} = 0,\\
\label{cov_amp-max}
&\partial_a F^{ab} = q\,N^b + J^b_{\rm ext}
\end{align}
with $\partial_a \equiv \partial/\partial x^a$ and $N^a$ the number $4$-current of the plasma electron fluid,
\begin{equation}
\label{number_current}
N^a(x) = \int \dot{x}^a f(x,\bm{v},\bm{a})\, \frac{d^3 v\, d^3 a}{1 + \bm{v}^2}
\end{equation}
where $\bm{v}^2 = v^\mu v_\mu$. The presence of the factor $1+\bm{v}^2$ in (\ref{number_current}) and the second term in (\ref{generalized_vlasov}) are related, as discussed in Ref. [\citenum{noble:2013}]. 

For clarity of presentation, the ions will be prescribed as a homogenous background and included in the external source $4$-current $J^a_{\rm ext}$.
\section{Fluid theory}
Kinetic theories are not always the most convenient tools for analytical investigation of the collective dynamics of charged matter (whether or not they include the radiative self-force). Furthermore, extensive computational resources are usually required to numerically solve the integro-differential systems of equations found in such theories.  

In practice, the $1$-particle distribution $f$ will usually contain more information than is needed and one may replace the Vlasov equation with a fluid theory that encodes $f$ using a subset of its velocity-acceleration moments. In general, macroscopic fluid theories are more analytically amenable and less computationally demanding than their kinetic counterparts, and their relationship with experiment is more immediate. 

Using the notation first introduced in Ref.~[\citenum{noble:2011}], the {\it natural} moments of $f$ are tensor fields on spacetime defined as 
\begin{equation}
S^{a_1\dots a_\ell : b_1\dots b_n}(x) = \int \dot{x}^{a_1}\dots \dot{x}^{a_\ell} \ddot{x}^{b_1}\dots \ddot{x}^{b_n} f(x,\bm{v},\bm{a}) \frac{d^3 v\, d^3 a}{1 + \bm{v}^2}.
\end{equation}
Indices associated with velocity are located to the left of the colon in $S^{a_1\dots a_\ell : b_1\dots b_n}$, whereas those associated with acceleration are located to the right of the colon. An absence of indices will be denoted by $\emptyset$ as follows:
\begin{align}
&S^\emptyset = \int f\,\frac{d^3 v\, d^3 a}{1+\bm{v}^2},\\
&S^{a_1\dots a_\ell : \emptyset} = \int \dot{x}^{a_1}\dots \dot{x}^{a_\ell} f\, \frac{d^3 v\, d^3 a}{1+\bm{v}^2},\\
&S^{\emptyset : b_1\dots b_n} = \int \ddot{x}^{b_1}\dots \ddot{x}^{b_n} f\, \frac{d^3 v\, d^3 a}{1+\bm{v}^2}.
\end{align}
Natural moments with an immediate physical interpretation include the number $4$-current $N^a = S^{a:\emptyset}$ and stress-energy-momentum tensor $T^{ab} = m S^{ab:\emptyset}$ of the plasma electron fluid. The scalar field $S^\emptyset$ is the relativistic enthalpy.

The Vlasov equation (\ref{generalized_vlasov}) may be recast as the infinite hierarchy of tensor equations
\begin{align}
\label{pde_10}
&\partial_{a} S^{a:\emptyset} = 0,\\
\label{pde_20}
&\partial_{a} S^{ab:\emptyset} - S^{\emptyset:b} = 0,\\
\label{pde_11}
&\partial_{a} S^{a:b} - S^{b:c}{ }_c - \tau^{-1} \bigg( S^{\emptyset:b} + \frac{q}{m} F^{b}{ }_c S^{c:\emptyset} \bigg) = 0,\\
\label{pde_30}
&\partial_{a} S^{abc:\emptyset} - S^{b:c} - S^{c:b} = 0,\\
\label{pde_21}
&\partial_{a} S^{ab:c} - S^{\emptyset:bc} - S^{bc:d}{ }_d - \tau^{-1} \bigg( S^{b:c} + \frac{q}{m} F^c{ }_d\,S^{bd:\emptyset} \bigg) = 0,\\
\label{pde_12}
&\partial_a S^{a:bc} - S^{b:cd}{ }_d - S^{c:bd}{ }_d - \tau^{-1} \left( 2 S^{\emptyset : bc} + \frac{q}{m} F^b{ }_d\,S^{d:c} + \frac{q}{m} F^c{ }_d\,S^{d:b} \right) = 0,\\
\label{pde_more}
&\dots
\end{align}
on spacetime, where $\dots$ indicates equations whose first term $\partial_{a_1} S^{a_1\dots a_\ell : b_1\dots b_n}$ satisfies $\ell + n > 3$. Furthermore, the identities $\ddot{x}^a \dot{x}_a = 0$ and $\dot{x}^a \dot{x}_a = -1$ lead to the constraints
\begin{align}
\label{con_20}
&S^{a}{ }_{a}{ }^{:\emptyset} = - S^\emptyset,\\
\label{con_11}
&S^{a:}{ }_{a} = 0,\\
\label{con_30}
&S^{ab}{ }_b{ }^{:\emptyset}= - S^{a:\emptyset},\\
\label{con_21_a}
&S^a{ }_a{ }^{:b} = - S^{\emptyset:b},\\
\label{con_21_b}
&S^{ab:}{ }_{a} = 0,\\
\label{con_12}
&S^{a:}{ }_{ab} = 0,\\
\label{con_more}
&\dots
\end{align}
where $\dots$ indicates equations containing natural moments with rank greater than $3$.

For practical purposes, a finite set of equations must be chosen from the infinite sequences (\ref{pde_10}-\ref{pde_more}), (\ref{con_20}-\ref{con_more}) and this may be achieved by introducing the bulk velocity $U^a$, bulk acceleration $A^a$
\begin{align}
\label{bulk_vel_def}
&U^a = S^{a:\emptyset}/S^\emptyset,\\
\label{bulk_accel_def}
&A^a = S^{\emptyset:a}/S^\emptyset
\end{align}
and the {\it centred} moments
\begin{align}
\notag
&R^{a_1 \dots a_\ell : b_1 \dots b_n}(x) = \int \bigg( \dot{x}^{a_1} - U^{a_1}(x) \bigg) \dots \bigg( \dot{x}^{a_\ell} - U^{a_\ell}(x)\bigg)\\
& \;\;\;\;\times \bigg( \ddot{x}^{b_1} - A^{b_1}(x) \bigg) \dots \bigg( \ddot{x}^{b_n} - A^{b_n}(x) \bigg)
 f(x, \bm{v}, \bm{a})\,\frac{ d^3 a\, d^3 v}{1+\bm{v}^2}
\end{align}
followed by the assumption that all centred moments of a particular rank or greater are negligible.

The rank $1$ centred moments $R^{a : \emptyset}$, $R^{\emptyset : a}$ trivially vanish due to the definitions of $U^a$, $A^a$, $S^{a:\emptyset}$, $S^{\emptyset:a}$. Setting all centred moments of rank $2$ or greater to zero is equivalent to demanding that the electron distribution has zero spread in velocity and acceleration, and this assumption collapses (\ref{pde_10}-\ref{pde_more}), (\ref{con_20}-\ref{con_more}) to the LAD equation for $U^a$ (see Ref. [\citenum{noble:2011}] for details).

A fluid modelling a collection of electrons whose distribution has a small, but non-negligible, spread about the bulk velocity and bulk acceleration is more subtle to construct. Following an approach analogous to the scheme introduced by Amendt~\cite{amendt:1986} for fluids without the radiative self-force, we introduce the scalar field $\epsilon = \sqrt{1+U^a U_a}$ and hypothesize that $R^{a_1 \dots a_\ell : b_1 \dots b_n} = {\cal O}(\epsilon^{\ell+n})$ with $S^\emptyset = {\cal O}(\epsilon^0)$, $U^a = {\cal O}(\epsilon^0)$, $A^a = {\cal O}(\epsilon^0)$. Thus, $R^{ab : \emptyset} = {\cal O}(\epsilon^2)$, $R^{a : b} = {\cal O}(\epsilon^2)$, $R^{\emptyset : ab} = {\cal O}(\epsilon^2)$ and inspection reveals that the total number of independent components of (\ref{pde_10}-\ref{pde_12}) matches the total number of independent components of the variables $S^\emptyset$, $U^a$, $A^a$, $R^{ab:\emptyset}$, $R^{a:b}$, $R^{\emptyset:ab}$ with all ${\cal O}(\epsilon^3)$ terms in (\ref{pde_10}-\ref{pde_12}) set to zero. However, in general, it is not possible to find solutions to the resulting field equations that also satisfy the constraints (\ref{con_20}-\ref{con_12}) with all ${\cal O}(\epsilon^3)$ terms set to zero in those constraints. Instead, inspired by previous work on fluids without the radiative self-force~\cite{weitzner:1987}, we impose the weaker condition that (\ref{con_20}-\ref{con_12}) need only be satisfied to ${\cal O}(\epsilon^3)$, which leads to 
\begin{align}
\label{rem_con_20}
& R^a{ }_a{ }^{:\emptyset} + S^\emptyset (1 + U^a U_a) = {\cal O}(\epsilon^3),\\
\label{rem_con_11}
& R^{a : }{ }_a + S^\emptyset U^a A_a = {\cal O}(\epsilon^3),\\
\label{rem_con_30}
& U^b R^a{ }_b{ }^{:\emptyset} = {\cal O}(\epsilon^3),\\
\label{rem_con_21_a}
& U^a R_a{ }^{: b} = {\cal O}(\epsilon^3),\\
\label{rem_con_21_b}
& U^a R^{b :}{ }_a + A_a R^{ab:\emptyset} = {\cal O}(\epsilon^3),\\
\label{rem_con_12}
& U^a R^{\emptyset:}{ }_{ab} + A_a R^{a:}{ }_b = {\cal O}(\epsilon^3).
\end{align}
A {\it warm} fluid model including the radiative self-force is obtained by setting to zero all terms that are ${\cal O}(\epsilon^3)$ in (\ref{pde_10}-\ref{pde_12}), and solutions to the resulting system of PDEs are sought that satisfy (\ref{rem_con_20}-\ref{rem_con_12}). 
\section{Equilibrium states in a magnetized plasma}
An investigation of (\ref{pde_10}-\ref{pde_12}), with all ${\cal O}(\epsilon^3)$ terms set to zero, leads to
\begin{align}
\label{zero_A}
&A^a = 0,\\
\label{zero_E}
&F^a{ }_b U^b = 0,\\
\label{eqm_R_eq_1}
&R^{\emptyset : bc} + \frac{1}{\tau}\bigg(R^{b:c} + \frac{q}{m}F^c{ }_d R^{bd : \emptyset}\bigg) = 0,\\
\label{eqm_R_eq_2}
&R^{b:c} + \frac{q}{2m}(F^c{ }_d R^{bd : \emptyset} - F^b{ }_d R^{cd : \emptyset}) = 0,\\
\label{eqm_R_eq_3}
&R^{\emptyset:bc} + \frac{q}{2m}(R^{d:b} F^c{ }_d +  R^{d:c} F^b{ }_d) = 0
\end{align}
when all of the fields are constant and, as expected for equilibrium states, (\ref{zero_A}) shows that the bulk acceleration vanishes and the electric field vanishes in the fluid frame as indicated by ({\ref{zero_E}). As expected, the above do not constrain the value of the magnetic field.

The centred moments $ R^{a:b}$, $R^{\emptyset:ab}$ may be eliminated from (\ref{eqm_R_eq_1}-\ref{eqm_R_eq_3}) to give
\begin{equation}
\label{matrix_FR}
\frac{q}{2m}\big\{{\cal F},\big\{{\cal F},{\cal R}\big\}\big\} - \frac{1}{\tau}\big[{\cal F},{\cal R}\big] = 0
\end{equation}
where $\big\{{\cal A}, {\cal B}\big\}$ denotes the anti-commutator and $\big[{\cal A},{\cal B}\big]$ denotes the commutator of the matrices ${\cal A}, {\cal B}$ with
\begin{align}
&{\cal R} = \big(R^a{ }_b{ }^{:\emptyset}\big),\\
&{\cal F} = \big(F^a{ }_b\big).
\end{align}
An elegant method for solving (\ref{matrix_FR}) exploits the algebraic properties of the electromagnetic field that arise because the electric field vanishes in the fluid frame. It may be shown that ${\cal F}{\cal F}^{\,\star} = {\cal F}^{\,\star}{\cal F} = 0$ where the matrix ${\cal F}^{\,\star} = \big(\varepsilon^a{ }_{bcd}\, F^{cd}/2\big)$ is the electromagnetic dual of ${\cal F}$, with $\varepsilon_{abcd}$ the Levi-Civita alternative symbol, and furthermore
\begin{align}
& {\cal P}{\cal P} = {\cal P},\\
& \check{{\cal P}}\check{{\cal P}}  = \check{{\cal P}},\\
& {\cal P}\check{{\cal P}} = \check{{\cal P}}{\cal P} = 0,\\
& {\cal P}+\check{{\cal P}} = {\cal I}
\end{align}
where ${\cal I}$ is the unit matrix and
\begin{align}
&{\cal P} = \frac{2{\cal F}^2}{{\rm tr}({\cal F}^2)},\\
&\check{{\cal P}} = \frac{2{\cal F}^{\,\star 2}}{{\rm tr}({\cal F^{\, \star}}^2)}.
\end{align}
Hence ${\cal F}$, ${\cal F}^\star$ naturally induce a pair of orthogonal idempotent matrices ${\cal P}, \check{{\cal P}}$ which can be used to decompose the tangent space on spacetime. The idempotent ${\cal P}$ projects vectors into the 2-dimensional subspace of the tangent space orthogonal to the bulk velocity $U^a$ and the magnetic field lines in the fluid frame.

The stress-energy-momentum tensor of the electron fluid is $T^{ab} = m S^{ab : \emptyset}$, and the heat flux is chosen to vanish in the equilibrium state. Since $S^{ab : \emptyset} = S^\emptyset U^a U^b + R^{ab : \emptyset}$, the choice
\begin{equation}
\label{zero_rem_con_30}
U_a R^{ab : \emptyset} = 0
\end{equation}
ensures that the heat flux vanishes and (\ref{rem_con_30}) is satisfied. It is then straightforward to use (\ref{zero_A} - \ref{eqm_R_eq_3}, \ref{zero_rem_con_30}) to show that the left-hand side of the constraints (\ref{rem_con_11}, \ref{rem_con_21_a} - \ref{rem_con_12}) vanish.

Equations (\ref{zero_E}, \ref{zero_rem_con_30}) suggest the introduction of the idempotent $\Pi = \big(\delta^a_b - U^a U_b / |U|^2\big)$, which projects vectors into the 3-dimensional subspace of the tangent space orthogonal to $U^a$, where $|U| = \sqrt{-U^a U_a}$. Since $R^{ab : \emptyset} = R^{ba : \emptyset}$ it follows $\Pi {\cal R} = {\cal R} \Pi = {\cal R}$ and, furthermore, $\Pi {\cal F} = {\cal F} \Pi = {\cal F}$ since the electric field vanishes in the fluid frame and $F_{ab} = - F_{ba}$. Hence, natural idempotents for constructing the solution to (\ref{matrix_FR}) are $\Pi {\cal P} \Pi$ and $\Pi \check{{\cal P}} \Pi$, where the former is simply ${\cal P}$, since $\Pi {\cal P} \Pi = {\cal P}$, and the latter projects vectors into the 1-dimensional subspace parallel to the magnetic field lines in the fluid frame.

A suitable form for ${\cal R}$ isotropic around the magnetic field lines is
\begin{equation}
\label{R_ansatz}
{\cal R} = \frac{p_\perp}{m} \Pi {\cal P} \Pi + \frac{p_\parallel}{m} \Pi \check{{\cal P}} \Pi 
\end{equation}
where the scalar $p_\perp$ (resp. $p_\parallel$) is the pressure transverse (resp. longitudinal) to the magnetic field lines. Hence, using (\ref{R_ansatz}) and the above properties, it follows $[{\cal F},{\cal R}] = 0$ whereas $\{{\cal F},\{{\cal F},{\cal R}\}\} \neq 0$ in general. By inspecting (\ref{matrix_FR}), we conclude that $p_\perp = 0$ and $p_\parallel$ can be chosen freely. It follows that the thermal motion is along the magnetic field lines only:
\begin{equation}
\label{eqm_R_final_form}
{\cal R} = \frac{p_\parallel}{m}\, \Pi \check{{\cal P}} \Pi.  
\end{equation}

The result (\ref{eqm_R_final_form}) has a straightforward physical explanation. An electron spiraling around the magnetic field lines will emit cyclotron radiation, whereas there is no emission due to inertial motion along the magnetic field lines. Hence, angular motion around the magnetic field lines will be damped by radiation friction and the electron's velocity will tend towards a constant vector parallel to the magnetic field lines.

The background ion number 4-current is $N^a_{\rm ion} = n_{\rm ion} \delta^a_0$ where the constant $n_{\rm ion}$ is the proper number density of the ions, and $S^{a : \emptyset} = N^a = N^a_{\rm ion}$ follows from the Maxwell equation (\ref{cov_amp-max}) since the electromagnetic field is constant and $J^a_{\rm ext} = - q N^a_{\rm ion}$. The enthalpy $S^\emptyset$ and bulk velocity $U^a$ are determined in terms of $p_\parallel$, $m$, $n_{\rm ion}$ using (\ref{bulk_vel_def}, \ref{rem_con_20}, \ref{eqm_R_final_form}) and the property $p_\parallel = {\cal O}(\epsilon^2)$. The latter follows from (\ref{eqm_R_final_form}) and the assumption $R^{ab : \emptyset} = {\cal O}(\epsilon^2)$.
\section{Electric waves}
A small amplitude, spatially harmonic, perturbation to the equilibrium proper number density, directed along the magnetic field lines, induces an electric wave that propagates along the magnetic field lines. In the following the $z=x^3$ axis is chosen to lie parallel to the magnetic field lines in the rest frame of the ions. We denote $t=x^0$ and introduce the proper electron number density $n = n_{\rm ion} + \delta n$ with $\delta n \propto \Re[ \exp(i k z - i \omega t)]$. Similar expressions are also introduced for the other field variables.
 
A linearization of (\ref{pde_10}-\ref{pde_12}), with all ${\cal O}(\epsilon^3)$ terms set to zero, around one of the equilibria introduced in the previous section yields a dispersion relation $\omega = \omega(k)$ solved by 
\begin{equation}
\label{dispersion}
\omega = \omega_p + \bigg(\frac{3}{2} \frac{k^2}{\omega_p} - \frac{3}{4} \omega_p \bigg) \theta - \frac{i \tau}{2}[\omega_p^2 - (2 k^2 + \omega_p^2)\theta] + {\cal O}(\tau^2,\theta^2)
\end{equation}
where the normalized equilibrium temperature $\theta = p_{\parallel}/(n_{\rm ion} m)$ and the plasma frequency $\omega_p = \sqrt{n_{\rm ion} q^2/m}$. The solution (\ref{dispersion}) was chosen because it reduces to $\omega=\omega_p$ in the limit $\tau\rightarrow 0$ and $\theta\rightarrow 0$. The dispersion relation also possesses unphysical solutions that diverge as $\tau\rightarrow 0$, which must be discarded. As required, the linearized solution to (\ref{pde_10}-\ref{pde_12}) can also be shown to satisfy the constraints (\ref{rem_con_20} - \ref{rem_con_12}).

The limit $\tau \rightarrow 0$ of  (\ref{dispersion}) yields a solution analogous to that of Clemmow and Willson's~\cite{clemmow:1956} relativistic generalization of the Bohm-Gross dispersion relation. The only difference is the numerical factor $3/4$ in the relativistic correction to the frequency shift in (\ref{dispersion}), which arises because of the anisotropy of $R^{ab : \emptyset}$ in equilibrium. The first damping term in (\ref{dispersion}) is an old result~\cite{burman:1969} common to kinetic theories with radiation reaction. The second damping term depends on the temperature and agrees with recent results presented elsewhere~\cite{noble:2013} as we will now show.

The kinetic theory developed in Ref.~[\citenum{noble:2013}] requires that the distribution $f$ has the form
\begin{equation}
\label{f_g_ansatz}
f(x,\bm{v},\bm{a}) = \sqrt{1+\bm{v}^2} g(x,\bm{v}) \delta^{(3)}\big(\bm{a}-\bm{A}(x,\bm{v})\big)
\end{equation}
for the LL kinetic theory to be recovered, where $g(x,\bm{v})$ is identified as the usual $1$-particle distribution on event-velocity space and $\delta^{(3)}$ is the $3$-dimensional Dirac delta. In Ref. [\citenum{noble:2013}], the conventional relativistic Vlasov equation was obtained for $g(x,\bm{v})$ in the limit $\tau\rightarrow 0$ and a dispersion relation for electric waves {\it without an external background field} was derived in terms of the equilibrium distribution $\widehat{g}(\bm{v})$. Specializing that dispersion relation to a spatially homogeneous and static distribution with zero transverse spread in velocity yields 
\begin{equation}
\label{kinetic_dispersion}
1 = \frac{q^2}{m}\int\limits\frac{h(u)}{\Psi^2\,(1 + i\tau\Psi)}\frac{du}{\sqrt{1+u^2}}
\end{equation}
where $\Psi = \omega\sqrt{1+u^2} - k u$, and the equilibrium distribution has the form $\widehat{g}(v^1, v^2, u) =  h(u)\delta(v^1)\delta(v^2)$ where $\delta$ is the $1$-dimensional Dirac delta. The judiciously chosen pre-factor $\sqrt{1+\bm{v}^2}$ in (\ref{f_g_ansatz}) ensures that the factor $1/\sqrt{1+u^2}$ appears in the measure in (\ref{kinetic_dispersion}).  

The proper number density $n_{\rm ion}$ of the ions and the normalized equilibrium temperature $\theta$ are related to $h$ in the usual way:
\begin{align}
&n_{\rm ion} = S^{0 : \emptyset}|_{\rm eqm} = \int \sqrt{1+u^2}\,h(u) \frac{du}{\sqrt{1+u^2}},\\
&\theta = \frac{S^{33 : \emptyset}|_{\rm eqm}}{n_{\rm ion}} = \frac{1}{n_{\rm ion}} \int u^2 h(u) \frac{du}{\sqrt{1+u^2}}
\end{align}
and the heat flux of the electron fluid vanishes in equilibrium
\begin{equation}
T^{3 0}|_{\rm eqm} = m\, S^{3 0 : \emptyset}|_{\rm eqm} = m \int u\sqrt{1+u^2}\, h(u) \frac{du}{\sqrt{1+u^2}} = 0.
\end{equation}
Hence
\begin{align}
\label{n_ion_h}
&n_{\rm ion} = \int h(u) du,\\
\label{theta_h}
&\theta =  \int u^2 h(u) du + \dots,\\
\label{zero_h}
&\int u\, h(u) du = 0
\end{align}
where $\dots$ indicates contributions that arise from ${\cal O}(u^3)$ integrands.

Using (\ref{n_ion_h}-\ref{zero_h}) and
\begin{equation}
\frac{1}{\Psi^2 (1+ i\tau\Psi)\sqrt{1+u^2}} = \frac{1}{\omega^2}\bigg(1 + \frac{2 k u}{\omega} - \frac{3}{2} u^2 + \frac{3 k^2 u^2}{\omega^2}\bigg) - i\tau \frac{1}{\omega}\bigg( 1 + \frac{ku}{\omega} + \frac{k^2 u^2}{\omega^2} - u^2\bigg) + {\cal O}(\tau^2, u^3) 
\end{equation}
it follows that (\ref{kinetic_dispersion}) yields the dispersion relation
\begin{equation}
\label{kinetic_dispersion_approx}
1 = \frac{\omega_p^2}{\omega^2}\bigg[1 - \frac{3}{2}\theta + \frac{3 k^2}{\omega^2}\theta - i \tau\omega\bigg(1 + \frac{k^2}{\omega^2}\theta - \theta\bigg)\bigg] + \dots 
\end{equation}
where $\dots$ indicates contributions that arise from ${\cal O}(u^3)$ terms in integrands or are ${\cal O}(\tau^2)$.

Terms that arise from ${\cal O}(u^3)$ integrands are negligible in the warm fluid approximation, and it may be shown (\ref{dispersion}) satisfies (\ref{kinetic_dispersion_approx}) when the terms indicated by $\dots$ are dropped from the calculation. Hence, the warm fluid theory presented here and the results of the kinetic theory developed in Ref. [\citenum{noble:2013}] agree, lending support to the methods established herein.
\section{Conclusion}
We have introduced a natural generalization of the standard warm fluid approximation that closes the hierarchy of velocity-acceleration moments generated from (\ref{generalized_vlasov}). Our model yields a relativistic generalization of the Bohm-Gross dispersion relation for electric waves propagating parallel to an ambient constant magnetic field in a magnetized plasma. Although the calculation in Ref. [\citenum{noble:2013}] assumes that the background electromagnetic field vanishes, the results found there are applicable to the present case if the equilibrium distribution models electrons confined to move freely along the background magnetic field lines.  The dispersion relation for electric waves found in Ref. [\citenum{noble:2013}] agrees with the present work.     

Inspection of  the imaginary part $\Im(\omega)$ of $\omega$ in (\ref{dispersion}) shows that plane waves with higher $k$ are damped less than those with lower $k$. The higher frequency components of a localized disturbance $\delta n$ in the electron number density $n$ are longer-lived than the shorter frequency components. Although $\delta n$ will spread due to the $k^2$ term in the real part $\Re(\omega)$ of $\omega$ in (\ref{dispersion}), the radiative self-force encourages additional spreading due to the relative enhancement of the higher frequency components.

Further ramifications of the warm fluid theory introduced here will be presented elsewhere. 

%
\section*{Acknowledgements}
The authors are members of the ALPHA-X consortium funded under EPSRC grant EP/J018171/1. DAB, AC and JG are supported by the Cockcroft Institute of Accelerator Science and Technology  (STFC grant ST/G008248/1).



\begin{thebibliography}{9}
\bibitem{eli} http://www.extreme-light-infrastructure.eu/
\bibitem{dirac:1938} P.A.M. Dirac, Proc. Roy. Soc. A {\bf 167}, 148 (1938)
\bibitem{ferris:2011} M.R. Ferris and J. Gratus, J. Math. Phys. {\bf 52} 092902 (2011)
\bibitem{hammond:2010} R.T. Hammond, Phys. Rev. A {\bf 81} 062104 (2010)
\bibitem{tamburini:2011} M. Tamburini {\it et al.}, Nucl. Instrum. Methods A {\bf 653}, (1) 181 (2011)
\bibitem{lehmann:2012} G. Lehmann and K.H. Spatschek, Phys. Rev. E {\bf 85}, 056412 (2012)
\bibitem{noble:2013} A. Noble, D.A. Burton, J. Gratus and D.A. Jaroszynski,  J. Math. Phys. {\bf 54} (2013) (to appear).\\
See also arXiv:1210.5467
\bibitem{hakim:1968} R. Hakim and A. Mangeney, J. Math. Phys. {\bf 9} 116 (1968)
\bibitem{noble:2011} A. Noble, {\it et al.} Proc. SPIE {\bf 8079} 0L (2011)
\bibitem{amendt:1986} P. Amendt, Phys. Fluids {\bf 29} 5 1458 (1986)
\bibitem{weitzner:1987} H. Weitzner in {\it Relativistic Fluid Dynamics}, Lecture Notes in Mathematics {\bf 1385} (ed. A. Anile, Y. Choquet-Bruhat) p. 211, Springer-Verlag (1987)
\bibitem{clemmow:1956} P.C. Clemmow and A.J. Willson, Proc. R. Soc. A {\bf 237} 117 (1956)
\bibitem{burman:1969} R. Burman, Phys. Lett A {\bf 30} 431 (1969)
\end{thebibliography}

\end{document}